\documentclass[11pt]{article}
\usepackage{amsmath}
\usepackage{amsfonts}
\newcounter{mnotecount}[section]

\renewcommand{\themnotecount}{\thesection.\arabic{mnotecount}}

\newcommand{\mnote}[1]
{\protect{\stepcounter{mnotecount}}$^{\mbox{\footnotesize
$
\bullet$\themnotecount}}$ \marginpar{
\raggedright\tiny\em
$\!\!\!\!\!\!\,\bullet$\themnotecount: #1} }

\newcommand{\tr}{\operatorname{tr}}

\def\be{\begin{equation}}
\def\ee{\end{equation}}
\def\bea{\begin{eqnarray}}
\def\eea{\end{eqnarray}}

\begin{document}
\title{Avoiding closed timelike curves with a collapsing rotating null dust shell}
\author{Filipe C. Mena$^{1,2}$, Jos\'e Nat\'ario$^{2}$ and Paul Tod$^{3}$\\
{\small $^1$ Departamento de Matem\'atica,
Universidade do Minho,
4710-057 Braga, Portugal}\\
{\small $^2$ Departamento de Matem\'atica, Instituto Superior T\'ecnico, 1049-001 Lisboa, Portugal}\\
{\small $^3$ Mathematical Institute,
University of Oxford,
St Giles' 24-29, Oxford OX1 3LB, U.K.}}
\date{18th October 2007}
\maketitle
\begin{abstract}
We present an idealised model of gravitational collapse, describing a collapsing rotating cylindrical shell of null dust in flat space, with the metric of a spinning cosmic string as the exterior. We find that the shell bounces before closed timelike curves can be formed. 
Our results also suggest slightly different definitions for the mass and angular momentum of the string.
\end{abstract}
\section*{Introduction}

A stationary, cylindrically-symmetric rotating solution of the Einstein equations containing closed timelike curves (CTCs) was first published in \cite{vanStockum}. Its possible role as a time-machine was discussed by Tipler \cite{Tipler1} and Bonnor \cite{Bonnor1}. Other metrics with CTCs include those of G\"odel \cite{Godel}, Newman-Unti-Tamburino \cite{NUT63,Misner}, Gott \cite{Gott} and the rotating cosmic string \cite{Deser}. In \cite{Bonnor2, Bonnor3} Bonnor has emphasised the need for a proper understanding for the formation of CTCs.

Several attempts have been made to rule out the possibility of creating CTCs \cite{Tipler2, Hawking, Deser2}. This would follow from a proof of the strong cosmic censorship conjecture, in the following sense: the domain of dependence of an appropriate\footnote{The simplest condition is to take it compact.} initial surface, evolving under the Einstein equations with a reasonable matter content, is (by construction) globally hyperbolic, and hence contains no CTCs. If in addition it is (generically) inextendible, which is the usual formulation of strong cosmic censorship, then there are no CTCs at all. It would then follow that, if one took an initial surface with initial data set up so that CTCs {\em could} form, in fact they would not.

It would be desirable to have non-stationary models which could provide test beds for this idea. In this paper we present a simplified example, obtained by matching a Minkowski interior across a collapsing, rotating cylindrical shell of null dust with positive energy density to a spinning cosmic string exterior\footnote{A similar matching was done across a stationary cylindrical shell in \cite{Jensen-Soleng}; thin shell collapse in 2+1 dimensions was studied in \cite{Peleg-Steif}, \cite{Cris-Olea}.}. We find that the shell bounces before the critical radius at which CTCs would be formed can be reached.

We follow the conventions of \cite{Wald}. The solutions considered are four-dimensional but the $z$-coordinate is ignorable, and we will omit it from the calculations.
\section{Interior}
We take the flat Minkowski interior,
\[
g^- = - d\tau^2 + d\xi^2 + d\eta^2,
\]
which we shall match to the spinning cosmic string exterior along the hyperboloid
\begin{equation}
- \tau^2 + \xi^2 + \eta^2 = a^2. \label{hyperboloid}
\end{equation}
As is well known, this hyperboloid is ruled by two families of null geodesics.
We choose coordinates $(u,\psi)$ on the hyperboloid adapted to one of these families, given by
\[
\begin{cases}
\tau= u \\
\xi = a \cos \psi - u \sin \psi \\
\eta = a \sin \psi + u \cos \psi
\end{cases}
\]
The induced metric on the hyperboloid is given in these coordinates by
\begin{equation}
h^- = 2a \, du \, d\psi + (u^2 + a^2) \, d \psi^2.\label{induced-}
\end{equation}
The second fundamental form of this hyperboloid is known to be proportional to the metric,
 a fact we obtain below by an indirect route.
\section{Exterior}
The metric for a spinning cosmic string is \cite{Deser, Jensen-Soleng}
\[
g^+ = - (dt + m d\varphi)^2 + C^2dr^2 + r^2 d\varphi^2,
\]
with $m>0$, $C>0$. Notice that since $g^+_{\varphi\varphi}=r^2-m^2$, the trajectories of the $\partial/\partial\varphi$-Killing vector define closed null curves if $r=m$ and CTCs if $r<m$. The parameters $C$ and $m$ are conventionally related \cite{Jensen-Soleng} to the mass per unit length $\mu$ and angular momentum per unit length $J$ of the string through
\begin{align}
\label{mass}
& \mu=\frac{C-1}{4C}; \\
\label{angularmom}
& J = \frac{m}4,
\end{align}
so that for positive mass $C> 1$. 
Our calculation, based on Killing vectors in the Minkowski space interior to the shell, suggests a slightly different
 identification of  mass and angular momentum per unit length in terms of $C$ and $m$, and the use of the Killing vectors in the exterior would give again a slightly different identification\footnote{The stationary observers on the exterior can be shown to be moving with speed $v=\frac{C^2-1}{C^2+1}$ with respect to the stationary observers on the interior; see \cite{Poisson} for a similar discussion regarding collapsing spherical null shells.}.

We must select a timelike surface in this spacetime which is ruled by a family of null geodesics,
 to be the matching surface.
The geodesic Lagrangian is
\[
L = \frac12\left[-(\dot{t} + m\dot{\varphi})^2 + C^2 \dot{r}^2 + r^2 \dot{\varphi}^2 \right],
\]
and the equations for null geodesics are
\begin{align}
& \frac{\partial L}{\partial \dot{t}} = - E  \quad \Leftrightarrow \quad \dot{t} + m\dot{\varphi} = E; \label{geodesic1} \\
& \frac{\partial L}{\partial \dot{\varphi}} = K  \quad \Leftrightarrow \quad r^2 \dot{\varphi} = K + mE; \label{geodesic2} \\
& L = 0 \quad \quad \Leftrightarrow \quad C^2 r^2 \dot{r}^2 = E^2 r^2 - (K + mE)^2. \label{geodesic3}
\end{align}
for constants $E$ and $K$. 
From \eqref{geodesic3} it is clear that if $\dot\varphi\neq0$ then $r$ always has a turning point,
 where it reaches its minimum value, which is
\[
b = \frac{K}{E} + m,
\]
where we assume that the geodesic is future-directed ($E>0$) and rotating in the positive direction ($\dot{\varphi} > 0$). We use this to introduce the parameter $\lambda$ through
\begin{equation}
r = b \sec \lambda, \label{parameterization}
\end{equation}
which, by \eqref{geodesic3}, satisfies
\[
\dot{\lambda} = \frac{E \cos^2 \lambda }{bC},
\]
choosing the positive root. Equations \eqref{geodesic2} and \eqref{geodesic1} then yield
\[
\begin{cases}
\varphi = C \lambda + \psi \\
t = bC \tan \lambda - mC \lambda
\end{cases}
\]
We may regard these equations, together with \eqref{parameterization},
 as a transformation to a new set of coordinates $\{\lambda, b, \psi\}$, for which the metric becomes
\be\label{met3}
g^+ = 2bC(b-m) \sec^2 \lambda \, d\lambda \, d\psi +
 C^2 db^2 - 2mC\tan \lambda \, db \, d\psi + (b^2 \sec^2 \lambda - m^2) \, d\psi^2.
\ee
The surfaces of constant $b$ are ruled by a family of null geodesics,
 as desired; we will now match one of these surfaces to the hyperboloid \eqref{hyperboloid}.

\subsection{The matching surface}
On the exterior, the metric induced on a hypersurface $\{ b = \text{constant} \}$ is
\[
h^+ = 2bC(b-m) \sec^2 \lambda \, d\lambda \, d\psi + (b^2 \sec^2 \lambda - m^2) \, d\psi^2.
\]
Introducing the coordinate
\[
u = b \tan \lambda
\]
this becomes
\begin{equation}
h^+ = 2C(b-m) \, du \, d\psi + (u^2 + b^2 - m^2)\, d\psi^2. \label{induced+}
\end{equation}
Comparing \eqref{induced+} with \eqref{induced-}, we see that the metric on the two surfaces matches if
\begin{equation}
\begin{cases}
a = C(b-m) \\
a^2 = b^2 - m^2
\end{cases}
\text{ or equivalently } \quad
\begin{cases}
a = \frac{2C}{C^2 - 1} m \\
b = \frac{C^2 + 1}{C^2 - 1} m
\end{cases} \label{matching}
\end{equation}
Note in particular that the matching requires $b>m$, so that the shell
 bounces before closed causal curves are revealed in the exterior.
  Thus the spacetime has the property claimed in the introduction,
   provided it has a physically reasonable, distributional matter content on the shell.

\subsection{Second fundamental form}
To compute the distributional energy-momentum tensor of the shell,
we next need to compute the second fundamental form of the matching
surface from the two sides. From the exterior, this is the second
fundamental form of the hypersurface $\{ b = \text{constant} \}$.
Note from (\ref{met3}) that a unit normal co-vector is $n=n_\alpha
dx^\alpha = Cdb$. Therefore
\[
n^\alpha\partial_\alpha = \frac{m \sin \lambda \cos \lambda}{bC(b-m)} \frac{\partial}{\partial \lambda}
+ \frac1C \frac{\partial}{\partial b}.
\]
Using
\[
\pounds_{n} d\lambda = d(\iota(n)d\lambda) + \iota(n) d(d \lambda) =
\frac{m (\cos^2 \lambda - \sin^2 \lambda)}{bC(b-m)} \, d\lambda + (\dotsb) \, db
\]
(where $\iota$ stands for contraction and the coefficient of $db$ doesn't contribute to the second fundamental form), and similarly
\[
\pounds_{n} db = \pounds_{n} d\psi = 0,
\]
we find
\begin{align}
K^+ & = \frac12 (\pounds_n g^+)_{|_{b=\text{constant}}} = 2b \sec^2 \lambda \, d \lambda \, d\psi + \frac{b}{C} \left( \sec^2 \lambda + \frac{m}{b-m} \tan^2 \lambda \right) d\psi^2 \nonumber \\
& = 2 \, du \, d\psi + \left( \frac{u^2}{C(b-m)} + \frac{b}{C}\right) d\psi^2. \label{extrinsic}
\end{align}

\section{Consequences of Matching}
The second fundamental form of the matching surface from the
 interior can be most easily obtained from \eqref{extrinsic}: setting $C=1$, $m=0$ and $b=a$ we have
\[
K^- = 2 \, du \, d\psi + \left( \frac{u^2}{a} + a \right) d\psi^2 = \frac1a \, h^-.
\]
(As noted above, this is to be expected for a hyperboloid). The jump in the second fundamental form is therefore
\[
\kappa = K^+ - K^- = - \frac{m}{C} \, d\psi^2,
\]
where we have used the matching conditions \eqref{matching}. The Darmois-Israel
formalism \cite{Israel} yields the stress-energy tensor
\[
T = S \delta(s),
\]
where $\delta$ is a Dirac delta-function, $s$ is the proper
length along the spacelike geodesics orthogonal to the matching surface and
\[
S = - \frac1{8\pi} (\kappa - (\tr \kappa) h)
\]
where $h=h^-=h^+$.
It is easily seen that  $\tr \kappa = 0$, and hence
\[
S = \frac{m}{8 \pi C} \, d\psi^2.
\]
Setting $l_\alpha dx^\alpha = d\psi$, we have
 $l^\alpha\partial_\alpha = \frac1a \frac{\partial}{\partial u}$. The contravariant version of $S$ is then
\[
S^{\alpha\beta}\partial_\alpha \otimes\partial_\beta  = \frac{m}{8 \pi C a^2} \frac{\partial}{\partial u} \otimes \frac{\partial}{\partial u},
\]
implying that the matter content of the shell is a null dust with positive energy density moving along the null geodesics which rule the matching surface. Note that $u$ is an affine parameter and $d\tau/du=1$ along these geodesics. 

In the interior, we have
\[
s = a - \sqrt{\rho^2 - \tau^2},
\]
where $\rho^2=\xi^2+\eta^2$.
Using the identity
\[
\delta(f(\rho)) = \frac{\delta(\rho-\rho_0)}{|f'(\rho_0)|}
\]
for functions $f$ with a single simple zero at $\rho=\rho_0$, we find
\[
\delta(s)=\delta\left(\sqrt{\rho^2 - \tau^2}-a\right) = \frac{\sqrt{\rho^2 - \tau^2}}{\rho} \, \delta\left(\rho-\sqrt{\tau^2 + a^2}\right) = \frac{a}{\rho} \, \delta\left(\rho-\sqrt{\tau^2 + a^2}\right).
\]
Therefore
\[
T^{\alpha\beta}\partial_\alpha \otimes\partial_\beta  = \frac{m}{8 \pi C a \rho} \, \delta\left(\rho-\sqrt{\tau^2 + a^2}\right) \frac{\partial}{\partial u} \otimes \frac{\partial}{\partial u}.
\]
Note that from \eqref{matching} and \eqref{mass} the surface energy density $\sigma$ can be written as
\be\label{sig1}
\sigma = \frac{m}{8 \pi aC \rho} = \frac{C^2 - 1}{16 \pi C^2  \rho} = \frac{(C + 1)}{2C} \frac{\mu}{2\pi \rho}.
\ee
The conservation equation for the null dust composing the shell is equivalent to the constancy of $2\pi \rho\sigma$, which is true by (\ref{sig1}), and then this quantity can be interpreted as the mass per unit length (in the $z$-direction). This suggests the identification of
\[
\hat\mu=\frac{C^2-1}{8C^2 }=\frac{(C+1)}{2C}\mu
\]
as the definition of mass per unit length, rather than $\mu$. Note
that $\hat\mu/\mu \to 1$ as $C \to 1$.

In cylindrical coordinates $\{\tau,\rho,\varphi\}$ for the interior, we have
\[
\frac{\partial}{\partial u} =\frac{\partial}{\partial \tau} + A \frac{\partial}{\partial \rho} + B \frac{\partial}{\partial \varphi},
\]
where $A$ and $B$ can be obtained from the conditions that $\partial/\partial u$ should be
 null and orthogonal to the hyperboloid. One finds
\[
\frac{\partial}{\partial u} =\frac{\partial}{\partial \tau} +
 \frac{\tau}{\rho} \frac{\partial}{\partial \rho} + \frac{a}{\rho^2} \frac{\partial}{\partial \varphi},
\]
and hence
\[
- T^{\alpha \beta} \left( \frac{\partial}{\partial \tau}\right)_\alpha \left( \frac{\partial}{\partial \varphi}\right)_\beta =
 \rho^2 T^{\tau \varphi} = a \sigma \delta\left(\rho-\sqrt{\tau^2 + a^2}\right).
\]
Therefore from \eqref{sig1} and \eqref{angularmom} the surface angular momentum density $j$ is
\[
j = a \sigma= \frac{m}{8 \pi  C\rho} = \frac{1}{C}\frac{J}{2 \pi \rho}.
\]
This suggests the identification
\[
\hat J=\frac{m}{4C}=\frac{J}{C}
\]
as the definition of the string's angular momentum per unit length,
rather than $J$. Again, $\hat J/J \to 1$ as $C \to 1$.

\section*{Acknowledgements}
FCM thanks Dep.~Matem\'{a}tica, Inst.~Sup.~T\'{e}cnico for hospitality, FCT (Portugal) for grant SFRH/BPD/12137/2003 and CMAT, Univ.~Minho for support. JN was partially supported by FCT (Portugal) through the Program POCI 2010/FEDER and the grant POCI/MAT/58549/2004. PT thanks Dep.~Matem\'{a}tica, Inst.~Sup.~T\'{e}cnico and CMAT, Univ.~Minho for hospitality.


\begin{thebibliography}{}


\bibitem{Bonnor1} Bonnor W B, The rigidly rotating relativistic dust cylinder, {\em J. Phys. A: Math. Gen.} {\bf 13} (1980) 2121--2132

\bibitem{Bonnor2} Bonnor W B, An exact, asymptotically flat, vacuum solution of Einstein's equations with closed timelike curves, {\em Class. Quantum Grav.} {\bf 19} (2002) 5951--5957

\bibitem{Bonnor3} Bonnor W B, Closed timelike curves in general relativity, {\em Int. J. Mod. Phys. D} {\bf 12} (2003) 1705-1708

\bibitem{Cris-Olea} Cris\'ostomo J and Olea R,  Hamiltonian treatment of the gravitational collapse of thin shells, {\em Phys. Rev. D} {\bf 69} (2004) 104023  

\bibitem{Deser} Deser S, Jackiw R and 't Hooft G, Three-dimensional Einstein gravity: dynamics of flat space, {\em Annals of Physics} {\bf 152} (1984) 220--235

\bibitem{Deser2} Deser S, Jackiw R and 't Hooft G, Physical cosmic strings do not generate closed timelike curves, {\em Phys. Rev. Lett.} {\bf 68} (1992) 267--269

\bibitem{Godel} G\"odel K, An example of a new type of cosmological solution of Einstein's field equations of gravitation, {\em Rev. Mod. Phys.} {\bf 21} (1949) 447--450

\bibitem{Gott} Gott J R, Closed timelike curves produced by pairs of moving cosmic strings: Exact solutions, {\em Phys. Rev. Lett.} {\bf 66} (1991) 1126--1129

\bibitem{Hawking} Hawking S W, Chronology protection conjecture, {\em Phys. Rev. D} {\bf 46} (1992) 603--611

\bibitem{Israel} Israel W, Singular hypersurfaces and thin shells in general relativity, {\em Nuovo Cimento} {\bf 44B} (1966) 1--14

\bibitem{Jensen-Soleng} Jensen B and Soleng H H, General-relativistic model of a spinning cosmic string, {\em Phys. Rev. D} {\bf 45} (1992) 3528--3533

\bibitem{Misner} Misner C W, Taub-NUT space as a counterexample to almost anything, {\em Relativity Theory and Astrophysics I: Relativity and Cosmology} ed. J Elhers,
Lectures in Applied Mathematics {\bf 8} 160--169 (American Mathematical Society, 1967)

\bibitem{NUT63} Newman E, Tamburino L \& Unti T, Empty-space generalization of the Schwarzschild metric, {\em J. Math. Phys.} {\bf 4} (1963), 915--923

\bibitem{Peleg-Steif} Peleg Y and Steif A, Phase transition for gravitationally collapsing dust shells in 2+1 dimensions, {\em Phys. Rev. D} {\bf 51} (1995) R3992--R3996

\bibitem{Poisson} Poisson E, A reformulation of the Barrab\`es-Israel null-shell formulation,  {\em gr-qc/0207101}

\bibitem{Tipler1} Tipler F J, Rotating cylinders and the possibility of global causal violation, {\em Phys. Rev. D} {\bf 9} (1974) 2203--2206

\bibitem{Tipler2} Tipler F J, Causality violation in asymptotically flat space-times, {\em Phys. Rev. Lett.} {\bf 37} (1976) 879--882

\bibitem{vanStockum} van Stockum W J, The gravitational field of a distribution of particles rotating around an axis of symmetry, {\em Proc. Roy. Soc. Edin.} {\bf 57} (1937), 135--154

\bibitem{Wald} Wald R M, {\em General Relativity} (Chicago: University of Chicago Press, 1984)
\end{thebibliography}
\end{document}